\documentstyle[12pt,fleqn,cite,epsfig]{article}

\def \cf {{\cal {F}}}
\def \ca {{\cal {A}}}

\def\be{\begin{equation}}
\def\ee{\end{equation}}
\def\bea{\begin{eqnarray}}
\def\eea{\end{eqnarray}}

\begin{document}
\begin{titlepage}
\begin{center}
\hfill hep-th/0005093\\
\hfill IP/BBSR/2000-16\\
\hfill IITK-HEP-00-02\\
\vskip .2in

{\Large \bf On Charged Strings and their Networks}
\vskip .5in

{\bf Alok Kumar$^{1}$ and Sudipta Mukherji$^{2}$\\
\vskip .1in
{\em 1. Institute of Physics,\\
Bhubaneswar 751 005, INDIA\\email: kumar@iopb.res.in}\\
{\em 2. Physics Department,\\Indian Institute of Tenchnology,\\Kanpur 208
016, INDIA\\
email: mukherji@iitk.ac.in}}

\end{center}

\begin{center} {\bf ABSTRACT}
\end{center}
\begin{quotation}\noindent
\baselineskip 10pt
We investigate  properties of several 
string networks in $D < 10$ which carry 
electric currents as well as electrostatic charge densities. 
We show the electric-current conservations as well as 
the force-balance condition of the  string
tensions on 3-string junctions in these networks. 
We also show the consistency of the 
above string networks from their world-volume point of view
by comparing the world-volume energy-density with the induced worldsheet
energy density of the supergravity solution. 
Finally,  we present new charged macroscopic string solutions in 
type II theories in $D=8$ and discuss certain issues related to 
their network construction.

\end{quotation}
\vskip 2in
May 2000\\
\end{titlepage}
\vfill
\eject

String-Network type BPS states have been  analyzed in  the past few 
years in string theory (with or without branes), as well as in 
other quantum field theories \cite{js} - \cite{khhh}. 
On the other hand, superconducting 
strings have been studied in field theories 
as they are expected to play important role in the evolution of our
universe at early time. Construction of such configurations in string
theory requires coupling of macroscopic strings to electromagnetic fields. 
Electromagnetically charged macroscopic strings are  known to exist in
heterotic string theory for some time \cite{ashokesen}. 
In a recent paper \cite{kuma}, these were extended to the type IIB string
theories. Moreover, it was also shown that BPS networks of such objects
with 1/4 supersymmetry can also be constructed. 

In this paper we analyze various aspects of such electrically 
charged string networks. We first show that in the 
above construction, the electric-currents flowing through the strings
are conserved on the 3-string junctions. In the absence of 
complete supergravity  solutions for such networks, these are done by 
examining the current flowing through them far away from any of the
3-string junction. Thus, although a redistribution of the current 
flow as well those of electrostatic charges are expected to take 
place in the complete solution near the junction,
we observe that the electric-currents
 emerging far away from these junctions are conserved. 
We then argue the consistency of charged string  from 
the point of view of world volume theories as well. We discuss this
aspect for charged-string networks which are constructed by 
applying a Lorentz transformation  involving time and an internal 
direction. It has earlier been  pointed out 
that configurations obtained by applying above 
transformations represent genuinely charged string network. 
In particular, by comparing the world-volume and supergravity 
expression for the energy-densities, we show that indeed 
a fundamental (charged) string ends on a D-string to 
form a 3-string junction. 

Finally, we present a new charged string solution in 8-dimensional 
type II theories. These solutions are obtained by observing a 
mapping between the heterotic and type II supergravity actions,
after suitable truncations. Since type II classical solutions also map 
on to the corresponding solutions in the heterotic theory, the
BPS nature of type II solutions is guaranteed.  
However, it is possible that they 
preserve different set of supersymmetries than the ones obtained
by another mapping between the truncated type IIB and heterotic 
theory, which was used in \cite{kuma}. To show the difference between the 
solutions presented in this paper and that of \cite{kuma}, 
we notice that in the charged macroscopic string solutions of
\cite{ashokesen}
and \cite{kuma}, the charges are acquired by fields which are identified as
Kaluza-Klein (KK) gauge fields: ${G_{\mu i} \pm B_{\mu i}}$, for $i=1,2$,
representing the internal directions. In the
present case however, charges are assigned
within an $SL(3)$ multiplet of gauge fields. Since two such multiplets are
formed by combinations $(G_{\mu 1}, B^{NS}_{\mu 2}, B^{RR}_{\mu 2})$
and $(G_{\mu 2}, B^{NS}_{\mu 1}, B^{RR}_{\mu 1})$, the solutions 
presented in this paper are necessarily different from 
the ones in \cite{kuma}.  Furthermore, we discuss 
%how various charged 
%string can be put together to form three-string junction 
%by analyzing various 
charge and current conservations around these junctions.

We now start with the discussion of the electric current conservations.
Let us consider a charged string solution  in D-dimensions which
is given by a supergravity configuration
 with non-zero 2-form  $B^{a}_{\mu \nu}$ and 1-form  $A^I_{\mu}$. 
Only nonzero components of $B^{a}_{\mu \nu}(r)$ are $B^{a}_{0, (D-1)}(r)$
and that of gauge field $A^I_{\mu}$ are  $A^I_0(r)$ and
$A^I_{D-1}(r)$. Here $r$ denotes the radial coordinate in $(D-2)$
dimensions transverse to the string and the superscripts $a, I$ on
$B_{\mu\nu}$ and $A_\mu$ distinguish between various two-form and 
one-form fields respectively.
Let us first focus on the 2-form fields. The charges associated to the
fields are defined by 
\begin{equation}
Z^a = \int d\Omega \>{}^*H^a_{\Omega}.
\label{charge}
\end{equation}
Here $H^a = dB^a$ and $*$ denotes the Hodge dual. The integral above is
taken over the $(D-2)$ transverse directions of the string.

Now let us think of a 3-string junction with its 
three legs coupling to different
values of $B^a$'s, denoted as $B^a_1$, $B^a_2$ and $B^a_3$. 
The corresponding charges are denoted as $Z^a_1$, $Z^a_2$ and $Z^a_3$.
Then, using above identifications,  the charge conservation: $Z^a_{(1)} +
Z^a_{(2)} = 
Z^a_{(3)}$ implies
\begin{equation}
\int_{\infty} d\Omega_{(1)} \> {}^*H^a_{\Omega_{(1)}} + 
\int_{\infty} d\Omega_{(2)} \>{}^*H^a_{\Omega_{(2)}} 
 = \int_{\infty} d\Omega_{(3)} \>{}^*H^a_{\Omega_{(3)}},
\label{conserve}
\end{equation} 
where we have now introduced subscripts $\Omega_{(i)}$'s for the angular
variable in the transverse space. 
Also, the subscripts of the integrals imply that they 
are evaluated far away from the junction. 
%This is to take care of the fact that the 3 legs of the 
%junction are oriented along different directions. In other words,
%since the coordinates transverse to three strings are different, 
%the radial and angular coordinates for them are also different and
%can be labelled as $\Omega_{1, 2, 3}$'s. We can therefore
%show the charge conservation, provided one can show (\ref{conserve}). 
In other words, the charge-conservation condition of the above type 
follows by evaluating the expression (\ref{charge}) along any one 
of the string of a 3-string junction (far away from it) and 
then by sliding the large spherical surface through the junction to 
the other side while deforming it in a manner to surround the 
remaining two strings. 
%This can be seen more easily for a 
%string junction in $D=4$, where the surrounding surface is 
%a circle. 

Same statements are true for the 
case of 1-form charges as well, when one
considers the charges corresponding to non-zero $A^I_{(D-1)}$'s. 
The field-strengths corresponding to $A^I_{(D-1)}$ are given by
$F^I_{(D-1) r}$.
% We once again define the Hodge-dual of these 
%field-strengths as: ${}^*F^I_{0 \Omega}$. We note that, unlike the 
%case of 3-forms, now the Hodge-dual carries a space-time index $0$,
%in addition to $\Omega$. 
The corresponding charges are given as:
\begin{equation}
J^I_0 = \int d\Omega \> {}^*F^I_{0 \Omega}
\label{1formcharge}
\end{equation}
where we have now kept the `time'-index on the charge to differentiate
them with other changes defined below. 
The charge conservation (which finally 
amounts to the electric-current conservation) then implies the 
condition:
\begin{equation}
{J^I_0}_{(1)} + {J^I_0}_{(2)} = {J^I_0}_{(3)}
\label{1fcurrent}
\end{equation}
As in the case of 2-form charge conservation condition 
(\ref{conserve}), eqn. (\ref{1fcurrent}) now follows 
from the following equation:
\begin{equation}
\int_{\infty} d\Omega_{(1)} \> {}^*F^I_{0 \Omega_{(1)}} + 
\int_{\infty} d\Omega_{(2)} \> {}^*F^I_{0 \Omega_{(2)}} = 
\int_{\infty} d\Omega_{(3)} \> {}^*F^I_{0 \Omega_{(3)}}.
\label{conserve2}
\end{equation}

Finally we discuss the case of gauge charges associated with 
1-form components $A^I_{0}$'s. Now the nonzero field-strengths
are: $F^I_{0 r}$ and their Hodge-duals are given as:
${}^*F^I_{(D-1) \Omega}$. The corresponding charges are now given as:
\begin{equation}
q^I_{(D-1)} = \int d\Omega \> {}^*F^I_{(D-1) \Omega}.
\label{1formcharge2}
\end{equation}                

An important difference between two 1-form charges defined in 
eqns. (\ref{1formcharge}) and (\ref{1formcharge2}) is 
that the later one depends on the direction along 
which the string lies as this charge is being measured by
the value of a field strength at a large orthogonal
distance from the string. The corresponding consistency condition 
of the charged 3-string junction is the force balance condition,
depending on their orientations. We now discuss these
aspects for examples presented earlier in  \cite{kuma}. 

We start by analyzing the 3-string junctions of 
charged macroscopic strings in $D=9$, discussed in section-(3.1) of 
\cite{kuma}. These are parameterized by a single solution-generating
parameter $\alpha$. Moreover, its action can be identified in 
ten-dimensions simply as a Lorentz-transformation involving 
time-coordinate $x^0$  and and an internal 
direction: $x^9$. The consistency of 
the network of such charged-string solutions is already 
known\cite{kuma}.
Explicit solution for the electrically charged fundamental string 
~( henceforth $(1, 0)$ string ), appearing in the networks, 
is presented in section-3.1 of \cite{kuma}. We only write 
down the 1-form potentials:
\be
\hat{A}^1_t = {{C {\rm sinh}\alpha ~{\rm cosh}\alpha\over { 2 
(r^5 + C {\rm cosh}^2\alpha)}}}, \>\>\> \hat{A}^1_8 = 0, 
\label{a1}
\ee
\be
\hat{A}^2_t = 0, \>\>\>
\hat{A}^2_8 = {-{C {\rm sinh}\alpha}\over { 2 
(r^5 + C)}}. 
\label{a2}
\ee
As we notice, the above $(1, 0)$-string solution is characterized
by two gauge fields for this $D=9$ example. They come from KK 
reduction of the ten-dimensional metric and antisymmetric
tensor fields. Then the $SL(2, Z)$ duality in ten-dimensions generates,
one more gauge field identified as the one coming from the 
KK reduction of the ten dimensional RR sector antisymmetric tensor. 
Nonzero gauge field components in the 
final $(p, q)$-string solution can then be written as:
\be
\hat{A}^1_t = {{C \Delta_q^{1/2} {\rm sinh}\alpha {\rm cosh}\alpha}\over {
2 (r^5 + C  \Delta_q^{1/2}{\rm cosh}^2\alpha)}}, \>\>\> \hat{A}^1_8 = 0, 
\label{nname}
\ee
\be
\hat{A}^2_t = 0, \>\>\>
\hat{A}^2_{i 8} = {-{C {\rm sinh}\alpha}\over { 2 
(r^5 + C \Delta_q^{1/2})}}({M}^{-1}_0 )_{i j} 
q_j,\>\> i=1, 2,
\ee
where $i=1, 2$ stand for the $B_{\mu \nu}$'s in 
NS-NS and R-R sectors of type IIB
in ten dimension and $M$ is a $2\times 2$ matrix parameterizing 
$SL(2)/SO(2)$ moduli. In the above equations, $\Delta_q = q_i
(M_0^{-1})_{ij}q_j$. 
By denoting the currents originating from the metric and the 
antisymmetric tensors respectively as $J^1$ and $J^2$,  the electric
current in a $(p, q)$  string\cite{schwarz} is:
\begin{equation}
J^1 = 0, \>\>\>
J^2_{i} = -C{{\rm sinh}\alpha\over 2}({M}^{-1}_0 )_{i j} 
q_j,\>\> i=1, 2. 
\label{planarcurrent}
\end{equation}
We now observe that, apart from an
$O(d, d)$ factor ${{\rm sinh}\alpha/2}$, the above electric currents
are proportional to the 2-form charges of the strings. 
As a result, the electric-current conservations hold at the
junctions directly due to the conservation of 2-form charges. 

We now discuss the electric-current conservation in string networks
constructed by starting with the $(1, 0)$
charged-macroscopic strings which are
T-dual to the ones we mentioned in the last paragraph. 
An analysis of the supersymmetry condition in this case 
implies that such solutions are possible in $D=8$. One then 
has six gauge fields, two each from the KK components
of $G_{\mu \nu}$, $B^{NS}_{\mu \nu}$ and 
$B^{R}_{\mu \nu}$. We denote the corresponding 
electric-currents as 
$J^1_a, J^{2 NS}_a, J^{2 R}_a$, $a=1, 2$. Then the results of
section-(3.2) in \cite{kuma} imply the following values of
the electric-currents for the $(1, 0)$ string solution
(in a complex notation):
\begin{equation}
J^1_1 + i J^1_2 = {\Delta_q^{1/2}\over 2} 
{\rm sinh}\alpha  e^{i \theta},
\>\>\> J^{2 i}_1 + i J^{2 i}_2 = 0
\label{current1}
\end{equation}
where an extra superscript $i$ in $J^2$, in the second equation 
above, stands for $NS-NS$ and $RR$ components
and the parameter $\theta$ is an angular $O(2, 2)$ parameter,
identified as a spatial rotation among the compactified 
coordinates $x^8$ and $x^9$.
However the network construction requires this parameter to 
be identical to the one associated with an $SL(2, Z)$ transformation,
namely eqn.(16) of \cite{schwarz}, which generates a 
$(p, q)$-string solution from a $(1, 0)$ one. 

Then for 3-string junctions, 
nonzero electric-current along a $(p, q)$-string-prong is given by:
\begin{equation}
J^1_1 + i J^1_2 ={1\over 2} {\rm sinh}\alpha  e^{\phi/2} 
(p - q \tau)
\label{current2}
\end{equation}
We then once again see the electric-current conservations in the
networks of such strings, following directly from the 2-form
charge conservations. 

Unlike electric-currents, there is no direct way to examine 
the status of the consistency of (static) electric-charge
densities 
coming from the gauge field components $A^1_t$ in (\ref{nname}), 
as these are expected to be redistributed in the full 
supergravity solution near a 3-string junction. 
One way to  settle the issue is 
to analyze supergravity solutions of such 3-string junctions.
In the absence of these solutions, 
for the moment, we note that the string tension of the 
charged macroscopic $(p, q)$-strings discussed above 
in eqns.(\ref{a1}-\ref{planarcurrent}) and 
(\ref{current1}-\ref{current2}) are given by 
\begin{equation}
T_{p, q} = {\Delta_q^{1/ 2}} C ({\rm cosh}\alpha + 1)/2 . 
\label{tens}
\end{equation}
Since the $O(d, d)$ parameter is only an overall factor, 
the tension balance continues to hold for both kinds
of string-networks mentioned above. This is because the spatial
orientations of these strings in a network are identical to the one for
the neutral ones. 

In view of applications in our later analysis , we now write down the 
induced world-sheet energy momentum tensors corresponding to 
general charged F-string solution\cite{kuma}. They are 
given in terms of $O(d, d)$ parameters $\alpha, \beta$:
\begin{eqnarray}
T_{00} &=& C {\rm cosh}\alpha {\rm cosh}\beta,\nonumber \\
T_{11} &=& C,\nonumber \\
T_{01} &=& {C\over 2} ({\rm cosh}\alpha - {\rm cosh}\beta ).
\label{energy-mom}
\end{eqnarray}
To get the corresponding answer for the examples considered 
in (\ref{a1}-\ref{planarcurrent}) and (\ref{current1}-\ref{current2}),
we have to set $\beta =\pm \alpha$. Therefore, for these configurations,
the
$(00)$ component of world-sheet stress tensor reduces to
%As it will be useful for world volume analysis of these networks,
%we now evaluate the energy density of the constituent strings.  This 
%is given by the $(00)$ component of the two dimensional effective energy 
%momentum tensor of these $(p,q)$strings. Following \cite{dghr}, we find
\begin{equation}
(T_{00})_{p,q} = {1\over 2}C {\Delta_q}^{1\over 2} (1 + {\rm cosh}
2 \alpha ).
\label{energymom}
\end{equation}
This should not, however, be compared with the tension calculated in 
(\ref{tens}) as $(T_{00})_{p,q}$ receives contributions from 
string tension as well as
from the gauge fields associated with electrically charged strings.
However we note that for $\alpha=0$, the expression (\ref{energymom}) matches
with the string tension, as expected in the neutral case.

%\vskip .2in

%\noindent {\bf 3. World-Volume Analysis}

%\vskip .2in

We now discuss the consistency of the charged string solutions
from the point of view of the D-string world-sheet theory. In this
context, we consider the example of  the $D=9$ charged string 
networks formed by a Lorentz-boost, parameterized by the 
parameter $\alpha$ mentioned above. Then, a classical solution
representing a 3-string junction of charged macroscopic strings
in the world-sheet theory, is given by the application 
of the Lorentz-boost on the solution of \cite{dasgupta} and
can be written as in eqns.(\ref{soln.}) and (\ref{Phi}) below:
\begin{eqnarray}
A_0  &=& - g x^1 {\rm cosh}\alpha, \qquad \>
\Phi \equiv A_9  = - g x^1 {\rm sinh}\alpha, \>\>\>x^1 > 0 \nonumber \\
 &=& 0 \>\qquad \qquad \qquad   \qquad \qquad = 0, 
\qquad\>\>\>\>\>\>\>\>\>\>\>\>\>\>\>x^1
< 0, 
\label{soln.}
\end{eqnarray}
where $\Phi$ is the scalar coming from the dimensional reduction 
of the world-volume gauge field 
from $D=10$ to $D=9$. The choice of the Lorentz-transformation 
parameter `${\alpha}$' is fixed through the results in 
section-(3.1), in particular (3.13) of \cite{kuma}.

To maintain supersymmetry one has to 
excite one more world-volume 
field, identified as the coordinate representing 
the F-string. Following \cite{ccjm},\cite{dasgupta}, in this case we have
\begin{eqnarray}
X^8 & = -g x^1,\>\>\>\>\>x^1 > 0, \cr
        & = 0, \>\>\>\>\>x^1 <0.
\label{Phi}
\end{eqnarray}
This is a $1/2$ supersymmetric solution in the world-volume theory.
The supersymmetry condition for this solution is obtained from 
that of the neutral string by the above Lorentz-transformation.
 
We now evaluate the energy of this configuration to identify it with 
the expression for 
$T_{00}$ of the F-string given in (\ref{energymom}). 
In order to proceed, following \cite{ccjm}, we first write down the
expression of the Hamiltonian associated with the above configuration.
After evaluating the expressions, one gets:
\begin{equation}
H = {1\over 2}(1 + {\rm cosh}2\alpha) H_0,
\label{hamiltonian}
\end{equation}
where $H_0$ is the Hamiltonian associated with the neutral 
3-string junction of \cite{dasgupta}. The first term inside 
the bracket in the right hand side corresponds to the contribution of
$X^8$ to the classical action whereas the second term
is the combined contribution from $A_0$ and $X^9$. 
As a result, the energy expression 
is modified by a factor $(1 + {\rm cosh} 2\alpha)/2$,  which 
precisely coincides with the energy density  of the charged string 
given in (\ref{energymom}), when retricted to $(1,0)$-string. 
We have therefore given a
world-volume argument in favor of the existence of 
the 3-string junction solutions of charged macroscopic strings
by identifying the relevant variables in the two approaches, 
namely $T_{00}$ in (\ref{energymom}) and 
$H$ in (\ref{hamiltonian}). In the limit $\alpha =0$, one also 
reproduces: $H_0 = T_f x^{8}$, a result following from the analysis
of \cite{ccjm,dasgupta}, with $T_f$ being the F-string tension. 
In other words, in \cite{ccjm,dasgupta},
the world-volume energy was associated with the string tension of
a `spike'-configuration interpreted as an F-string. We 
observe that similar interpretation holds in the case of 
charged strings as well, provided one takes into account 
the contribution of the charges in the supergravity solution.

%It should also be possible to find out similar world-volume 
%arguments for the existence of 3-string junctions of 
%other charged macroscopic strings, such as the ones 
%discussed in (\ref{}) and in section-3.2 of \cite{kuma}
%using T-duality rotations on the world-volume. 

%\vskip .2in

%\noindent {\bf 4. New electrically charged string configurations and their
%networks}

%\vskip .2in
We now give explicit construction for some new charged string
configuration
in eight dimensions. We further discuss how various 
conservation laws are satisfied around the junction for
such strings.

Following \cite{luroy}, we start with 
a truncated version of eight dimensional type IIB 
action. In Einstein frame  the action can  be written as 
\begin{eqnarray}
S = \int d^8x [&&R - {1\over 2}\{ (\partial \sigma)^2
+ (\partial \phi_1)^2 + (\partial \phi_2)^2\}\nonumber \\
&&-{1\over{12}}\{e^{-\phi_1 +{1\over{\sqrt 3}}\phi_2}{{H_3}^{(1)}}^2 +
e^{\phi_1 + {1\over{\sqrt 3}}\phi_2}{{H_3}^{(2)}}^2 +
e^{- {2\over{\sqrt 3}}\phi_2}{{H_3}^{(3)}}^2 \}\nonumber \\
&& -{1\over{4}}e^{\sigma}\{e^{-\phi_1 - {1\over{\sqrt
3}}\phi_2}{{F_2}^{(1)}}^2 +
e^{-\phi_1 - {1\over{\sqrt 3}}\phi_2}{{F_2}^{(2)}}^2 +
e^{ {2\over{\sqrt 3}}\phi_2}{{F_2}^{(3)}}^2 \}\nonumber \\
&& - {1\over{4}}e^{-\sigma}\{e^{\phi_1 - {1\over{\sqrt
3}}\phi_2}{{\cf_2}^{(1)}}^2 +
e^{-\phi_1 - {1\over{\sqrt 3}}\phi_2}{{\cf_2}^{(2)}}^2 +
e^{ {2\over{\sqrt 3}}\phi_2}{{\cf_2}^{(3)}}^2 \}].
\label{one}
\end{eqnarray}

We would like to make few comments about the origin of different
fields in this action. The action contains three scalars $\sigma$,
$\phi_1$ and $\phi_2$. They are certain linear combinations of ten
dimensional dilaton, and the two scalars that originate due to
compactification from ten to eight dimensions. Three different three-form
field strengths are denoted above as $H_3^{(i)}$. Furthermore, there 
are six two-form field strengths. Out of them $F_2^{(i)}$ come from
reduction of various antisymmetric tensors in ten dimension. The other
set $\cf_2^{(i)}$ have their KK origin.  In order to keep our
discussion simple, we have set all the 
other fields to zero including the zero forms (axions) that appear in the
eight dimensional action.
Various details of eight dimensional type IIB supergravity action can
be found in \cite{luroy,cjlp}.  As discussed previously, type IIB string
in eight dimensions has $SL(3,R)$ symmetry. Defining 
$H_3 = dB_2, F_2 = dA_1$ and ${\cal{F}} = d {\cal{A}}_1$, it is easy to
see that (\ref{one}) is invariant under 
\begin{eqnarray}
&& g_{\mu\nu} \rightarrow g_{\mu\nu}, ~~\sigma
\rightarrow \sigma,\nonumber \\
&& {\cal {M}} \rightarrow {\bf\Lambda} {\cal{ M}} {\bf\Lambda}^T, 
~{\bf{\cal{A}}_1} \rightarrow
\bf \Lambda {\bf{{\cal{A}}_1}}, \nonumber \\
&&{\bf A}_1 \rightarrow {\bf\Lambda} {\bf A}_1, ~~{\bf B}_2 \rightarrow
({\bf\Lambda}
^{-1})^T {\bf B}_2,
\label{slfour}
\end{eqnarray}
where ${\bf\Lambda}$ is a global $SL(3,R)$ matrix. $\cal M$ is a 
matrix with diagonal entries
 $(e^{-\phi_1 + {1\over \sqrt
3}\phi_2},$ $e^{\phi_1 + {1\over{\sqrt 3}}\phi_2}, e^{- {2\over \sqrt
3}\phi_2})$. In (\ref{slfour}), ${\bf {\cal A}_1}$ is defined as
three dimensional column matrix with entries $\ca_1^{(1)}$,
 $\ca_1^{(2)}$ and $\ca_1^{(3)}$. We have also defined  ${\bf A}_1$ and
${\bf B}_2$ in a similar manner. In the following, we will be using only
$SO(3)$ subgroup  ${\bf\Lambda}$ of $SL(3,R)$. This can be represented by
Euler angles $\theta, \phi$ and $\psi$. 

The macroscopic string solution of this theory in Einstein frame
can be written down as 
\begin{eqnarray}
ds^2 = {1\over {[1 + N G(r)]^{2\over 3}}}
&&[-dt^2 + (dx^7)^2] + {q^2G(r)\over {4N [1 + N G(r)]^{5\over 3}}}
[-dt  + dx^7]^2 \nonumber \\
&& + [1 + N G(r)]^{1\over 3}(dr^2 + r^2 d\Omega_5^2 ),
\label{two}
\end{eqnarray}

with
\begin{eqnarray}
&&\phi_1 = -{1\over 2} {\rm log}[1 + N G(r)], \>\>\>
 \phi_2 = {1\over {2\sqrt 3}}{\rm log}[1 + N G(r)]\nonumber \\
&&{ B_{t7}}^{(1)} = -{NG(r)\over {[1+ NG(r)]}}, \>\>\>
 {\cal {A}}_t^{(2)} = -{\cal{A}}_7^{(2)} = {qG(r)\over {2[1+NG(r)]}},
\label{three}
\end{eqnarray}
All other fields are set to zero. In the above expressions,
$G(r) = {1\over {4 \omega_5 r^4}}, ~~N = M {\rm
cosh}^2{\delta\over 2},
~~q = M{\rm sinh}\delta$.
Here $\omega_5$ is the unit volume of the $5$-sphere.
The tension of the string is given by $T = N$.
%Clearly, the string is aligned along the $x^7$ direction and is charged
%under $\ca_t^{(2)} $ and $\ca_7^{(2)}$. 
Using the asymptotic
behavior of various fields, we find that the
NS-NS two form charge $(Z)$, the electric 
charge $(Q)$ and the electric current $(J)$ for the solution are given
respectively
by:
\begin{equation}
{Z = N, ~~Q = q, ~~{\rm and}~~J = q.}
\label{four}
\end{equation}
Notice that the electric charge and current are same for the solution.
One way to obtain this solution is to first embed eight dimensional
heterotic string theory in type IIB string theory. Then we can translate 
the  charged heterotic string solutions of \cite{ashokesen} in terms of 
type IIB variables.

Now, using the symmetry of eight dimensional type IIB
strings, one can construct an $SL(3, Z)$ multiplet of above 
string solution. 
We do not give this explicitly, since it is straightforward
to write them down. We directly write down the charges
${{\bf Z}^\prime}$ and  and current ${{\bf J}^\prime}$
that follow from the above configuration:
\begin{eqnarray}
{\bf Z}^\prime &=& \pmatrix{{Z^\prime}^{(1)}\cr
{Z^\prime}^{(2)}\cr
{Z^\prime}^{(3)}\cr} = N_{(z_1,z_2,z_3)}
\pmatrix{{\rm cos}\theta {\rm cos}\phi\cr
{\rm sin}\theta {\rm cos}\phi\cr
{\rm sin}\phi\cr},\nonumber \\
{{\bf J}^\prime} &=& \pmatrix{{J^\prime}^{(1)}\cr
{J^\prime}^{(2)}\cr
{J^\prime}^{(3)}\cr} = q_{(z_1,z_2,z_3)}
\pmatrix{-{\rm sin}\theta {\rm cos}\psi -
{\rm cos}\theta {\rm sin}\phi {\rm sin}\psi\cr
{\rm cos}\theta{\rm cos}\psi - {\rm sin}\theta {\rm sin}\phi {\rm
sin}\psi\cr
{\rm cos}\phi {\rm sin}\psi},
\label{seven}
\end{eqnarray}
where $N_{(z_1,z_2,z_3)} = {\sqrt{z_1^2 +z_2^2 + z_3^2}}N$
and  $q_{(z_1,z_2,z_3)} = {\sqrt{z_1^2 +z_2^2 + z_3^2}}q$.
Parameters $\theta, \phi, \psi$ in equation (\ref{seven}) are
the Euler angles, as mentioned earlier. 
We would now like to identify ${\bf Z}^\prime = (z_1,z_2,z_3)^T M 
{\rm cosh}^2\delta/2$. This, in turn, fixes  part of the 
$SO(3)$ group parameters $\theta$ and $\phi$. Namely,
\begin{eqnarray}
{\rm cos}\theta {\rm cos}\phi = { z_1\over
{\sum_{i=1}^3\sqrt{z_i^2}}},\>\>
{\rm sin}\theta {\rm cos}\phi = { z_2\over
{\sum_{i=1}^3\sqrt{z_i^2}}},\>\>
{\rm sin}\phi = { z_3\over
{\sum_{i=1}^3\sqrt{z_i^2}}}.
\label{fix}
\end{eqnarray}
Notice that in this way, $Z_{(1,0,0)}$ string 
corresponds to electrically charged F-string, 
~$Z_{(0,1,0)}$ is an electrically charged D-string, and,
~$Z_{(0,0,1)}$ is a ten dimensional D-3 brane wrapped on two internal
circles. The subscript on $Z$ 
in the last line denotes their $z$ quantum numbers. In a similar
manner, we can define the electric charge of the configuration in 
(\ref{seven})
as ${\bf J}^\prime = (q_1, q_2, q_3)^T M {\rm sinh}\delta$. However,
$q$'s are not independent quantities. Rather, they are determined by
$z$'s and one of the $SO(3)$ group parameter $\psi$. Explicitly,
using (\ref{fix}) ~in (\ref{seven}) ~for currents, we get
\begin{eqnarray}
\pmatrix{J^\prime_1\cr J^\prime_2\cr J^\prime_3\cr} =
\pmatrix{ {{z_2 {\sqrt{z_1^2+z_2^2+z_3^2}} {\rm cos}\psi
- z_1 z_3 {\rm sin}\psi}\over {\sqrt{z_1^2 + z_2^2}}}\cr
{ {z_1 {\sqrt{z_1^2+z_2^2+z_3^2}} {\rm cos}\psi
- z_2 z_3 {\rm sin}\psi}\over {\sqrt{z_1^2 + z_2^2}}}\cr
{\sqrt{z_1^2 + z_2^2}}{\rm sin}\psi }.
\label{qch}
\end{eqnarray}
%>From the ${\bf Z}^\prime$ and
%${\bf Q}^\prime$ triplets, it is interesting to note  that for a given
%${\bf Z}^\prime$ triplet
%(that is for fix $z_1,z_2,z_3$), 
Here we note that one can get different $\bf J^\prime$'s
for different value of $\psi$. 

In order to construct a junction configuration, one can consider 
a special class of solutions, namely where
a $(z_1,0,0)$ and  $(0,z_2,0)$ strings meet. From the ${\bf Z}$ charge 
conservation, we see that resulting string must be a  $(z_1,z_2,0)$
string. Furthermore, in order to analyze the stability of a junction
of three such strings we notice that their string tensions are
given by expressions: 
%we have also to check various force balance
%and conservations around the junction. Since these three different
%strings have different tensions, we should first check that the 
%tension is balanced around the junction. This, in turn, will
%fix the angles between the strings. In order to proceed,
%let us first notice that the tensions of the strings considered
%here depend upon the electric charges in a very minimal way, 
%namely,
\begin{equation}
T_{(z_1,z_2,z_3)} = {\sqrt{z_1^2 + z_2^2 + z_3^2}}T_{(1,0,0)}
{\rm cosh}^2{\delta\over 2},
\label{tb}
\end{equation}
where $T_{(1,0,0)}$ is the tension of electrically neutral
$(1,0,0)$ string, and, from (\ref{four}), we see that it is given
by $T_{(1,0,0)} = M$. 
%From earlier works on electrically 
%neutral string junctions \cite{dasgupta}, we know that
%the angles form between three 
%constituent strings depend on only on the charges $z_1, z_2,z_3$.
%Now since, in our case, $\delta$ does not mix with $z_1$ and $z_2$,
%the angles between various strings are going to be independent
%of $\delta$. 
Once again, since $\delta$ does not mix with $z_1$ and $z_2$,
 various angles between the strings in a network
would be same as their electrically neutral counterparts.
Beside $Z$ charge conservation and tension balance, our string
junction have to satisfy other  constraints as discussed before. 
One of them comes from electric current conservation.
%First, we will check the electric charge conservation.
%>From (\ref{qch}), we see that the currents for  $(z_1,0,0)$, 
%$(0,z_2,0)$ and $(z_1,z_2,0)$ strings are
%\begin{eqnarray}
%{{\bf J}^\prime}_{(z_1,0,0)} = \pmatrix{0\cr z_1{\rm cos}\gamma\cr
%z_1 {\rm sin} \gamma \cr},
%~~{{\bf J}^\prime}_{(0,z_2,0)} = \pmatrix{z_2 {\rm cos}\gamma \cr 0\cr
%z_2 {\rm sin} \gamma \cr},
%~~{{\bf J}^\prime}_{(z_1,z_2,0)} = \pmatrix{z_2 {\rm cos}\gamma \cr
%z_1 {\rm cos} \gamma \cr {\sqrt{z_1^2 + z_2^2}} {\rm sin}\gamma}.
%\label{jch}
%\end{eqnarray}
We notice that in general the electric charge is not conserved 
unless $\psi = 0$. Thus the only allowed charged string junction
in this class is for 
$\psi = 0$, when
\begin{eqnarray}
{{\bf J}^\prime}_{(z_1,z_2,0)} = {{\bf J}^\prime}_{(z_1,0,0)} +
{{\bf J}^\prime}_{(0,z_2,0)}.
\label{ccon}
\end{eqnarray}
This is the `Kirchoff's law' for the junction. It simply says
that the algebraic sum of the currents around the junction must be zero.
At this stage, the restriction on $\psi$ for charge
conservation might seem unnatural. However, we should notice
that we started with a very special class of solutions (\ref{two}). 
We thus believe that the restriction on $\psi$
is an artifact of restricting ourselves within this special 
class of configuration. We expect that such restrictions
can be avoided if we look for more general class 
of string junctions.
Now, turning back to (\ref{three}),  we see that the seed solution that we
started with
has ${\cal A}_t^{(2)} = -{\cal A}_7^{(2)}$. 
%This, in turn, says that
%the charge density is equal to current for the seed solution.
%Now, under  $SO(3)$ rotation (which is essentially used to generate
%the triplet in section 2)  ${\cal A}_t^{(2)}$ and ${\cal A}_7^{(2)}$
%transforms exactly in the same way (though we have discussed so far
%only the  ${\cal A}_7^{(2)}$ transformations in order to 
%analyze various charge densities). Thus after the $SO(3)$ rotation,
%the charges are given exactly by the right hand side of the
%equation (\ref{qch}). 
This in turn,  leads to a condition on the charge densities 
similar to the one in (\ref{ccon}) for the currents. 
%current conservation ( for
%$\gamma =0$) as 
%\begin{equation}
%{{\bf Q}^\prime}_{(z_1,z_2,0)} = {{\bf Q}^\prime}_{(z_1,0,0)} +
%{{\bf Q}^\prime}_{(0,z_2,0)}.
%\label{curcon}
%\end{equation}

Now, for charge string junctions satisfying the above
conservation and stability criteria, we can put them together 
to construct a string network as in \cite{asenone,kuma}. However,
unlike in the previous cases, in our case these conditions
only guarantee their classical stability properties. In addition,
as in \cite{asenone,kuma}, one has to examine 
supersymmetry property as well to find out if these are 
BPS string networks or not. 
In the later case, they will decay into 
other BPS states. It is of interest to examine if the
corresponding final states are again string-networks 
and whether they are built out of charged or neutral strings.
These statements can be made more precise by thinking of 
string networks on tori\cite{asenone,bkm}. Then the final 
mass formula for a `particle-like' object
will carry the overall factor ${\rm cosh}^2 {\delta\over 2}$ appearing in 
the string tension in (\ref{tb}) which has a minimum at $\delta=0$. 
One however needs a more careful study, including quantum 
corrections, to clarify this further.

We conclude by stating that 
one of our main motivation for studying charged, current carrying junction
configurations and their networks is to set a framework for understanding
entropy associated with the network when compactified on two-torus.
Electrically charged networks that we discuss in this paper can 
be viewed as excitations over neutral networks. These excitations, in
some examples, preserve a 
fraction of original supersymmetry. We believe (as was in the case of
identifying string states associated with black hole entropy; see for
example \cite{callan}, \cite{dabholkar}) that identification of the
degrees of freedom for such excitations will play important role in
understanding entropy associated with network on torus. We hope to return 
back to this issue in the future.

\vskip .2in

{\bf Acknowledgements}: We have benefitted from discussions with 
Sunil Mukhi and Ashoke Sen.

\vfil
\eject

%%%%%%%%%%%%%%%%%%%%%%%%%%%%%%%%%%%%%%%%%%%%%%%%%%%%%%%%%%%%%%%%%%
%\Newpage

\end{document}